\documentstyle[12pt]{article}
\begin{document}
\topskip 2cm
\begin{titlepage}

\begin{center} 
{\large{\bf Standard Solar Neutrinos and The Standard Solar Model}} \\ 
\vspace{0.5cm}
{\large Arnon Dar} \\ 
\vspace{0.5cm} {\sl Department of Physics}\\ 
{\sl Technion - Israel Institute of Technology}\\ 
{\sl Haifa 3200, Israel}\\
\vspace{1.0cm} 
\end{center} 

\begin{abstract} 

The standard solar model (SSM) yield a $^8$B solar neutrino flux which
is consistent within the theoretical and experimental uncertainties with
that observed at Super-Kamiokande. The combined results from the Super-
Kamiokande and the Chlorine solar neutrino experiments do not provide a
solid evidence for neutrino properties beyond the minimal standard
electroweak model. The results from the Gallium experiments and
independently the combined results from Super-Kamiokande and the Chlorine
experiment imply that the $^7$Be solar neutrino flux is strongly
suppressed compared with that predicted by the SSM. This conclusion,
however, is valid only if the neutrino absorption cross sections near
threshold in Gallium and Chlorine do not differ significantly from their
theoretical estimates. Such a departure has not been ruled out by the
Chromium source experiments in Gallium. Even if the $^7$Be solar neutrino
flux is suppressed compared with that predicted by the SSM, still it can
be due to astrophysical effects not included in the simplistic SSM. Such
effects include spatial and/or temporal variations in the temperature in
the solar core induced by the convective layer through g-modes or by
rotational mixing in the solar core, and dense plasma
effects which may strongly enhance p-capture by $^7$Be relative to
e-capture. The new generation of solar observations, which already look
non stop deep into the sun, like Super-Kamiokande through neutrinos, and
SOHO and GONG through acoustic waves, may be able to point at the correct
solution; astrophysical solutions if they detect unexpected temporal
and/or spatial behaviour, or particle physics solutions if
Super-Kamiokande detects characteristic spectral distortion or temporal
variations (e.g., the day-night effect) of the $^8$B solar neutrino flux . 
If Super-Kamiokande will discover neither spectral distortions nor time
dependence of the $^8$ solar neutrino flux then, only future solar
neutrino experiments such as SNO, BOREXINO and HELLAZ, will be able to
find out whether the solar neutrino problem is due to neutrino properties
beyond the minimal standard electroweak model or whether it is just a
problem of the too simplistic standard solar model. 
  
\end{abstract}

\end{titlepage}

\section{Introduction} 

Solar neutrinos have been detected on Earth in four pioneering solar
neutrino ($\nu_\odot$) experiments, in roughly the expected numbers,
demonstrating that the sun is indeed powered by fusion of hydrogen into
helium. However the precise counting rates in these experiments are 
significantly below (see, e.g., Gavrin, these proceedings, Suzuki 1997) 
those predicted by traditional Standard Solar Models (SSM) (see, e.g.,
Bahcal and Pinsonneault 1995; Castellani et al. 1997).
This discrepancy, which has persisted for 25 years, has become known as
the solar neutrino problem (SNP). It has attracted much attention of both
astrophysicists and particle physicists for two main reasons.  First,
astrophysicists were surprised to find out that the nuclear reaction rates
inside the sun deviate significantly from that predicted by the simple
SSM. Second, particle physicists found that natural extensions of the
minimal standard electroweak model (SEM) can solve elegantly the SNP.
However, when astronomers had a closer look at the sun through
helioseismology, X-ray and UV observations it turned out to be ``a
bewildering turmoil of complex phenomena'', showing unexpected features
and behavior at any scale. It has a strange complex internal rotation,
unexplained magnetic activity with unexplained 11 year cycle, unexpected
anomalies in its surface elemental abundances, unexplained explosions in
its atmosphere and unexplained mechanism that heats its million degree
corona and accelerates the solar wind.  Perhaps the surface of the sun is
complex because we can see it and the center of the sun is not only
because we cannot? Perhaps the SSM which has been improved continuously
over the past three decades (see e.g., Clayton 1968, Bahcall 1989), which
still assumes spherical symmetry, no mass loss or mass accretion, no
angular momentum loss or gain, no differential rotation and zero magnetic
field through the entire solar evolution, is a too simplistic picture and
does not provide a sufficiently accurate description of the core of the
sun and/or the neutrino producing reactions there? 

\noindent
Indeed, here I will show that the solar neutrino problem does
not provide solid evidence for neutrino properties beyond the SEM 
and that standard physics solutions to the SNP are still possible. In 
particular I will argue that:

\noindent
1. The $^8$B solar neutrino flux predicted by the standard solar model (SSM)
is consistent within the theoretical and experimental uncertainties with
that observed at Super-Kamiokande. 

\noindent
2. There is no solid evidence for a real deficit of $^7$Be solar neutrinos.

\noindent
3. A real deficit of $^7$Be solar neutrinos, if there is one, 
may still be explained by standard physics and/or astrophysics.

\noindent
4. Only direct observations of spectral distortions and/or flavor change of
$\nu_\odot$'s in future $\nu_\odot$ experiments, like Super-Kamiokande,
SNO, Borexino and HELLAZ may establish that neutrino properties beyond the
SEM are responsible for the SNP. 

\noindent
5. The three new solar ``observatories'', which are already running and
looking into the solar interior, the Super-Kamiokande solar neutrino
observatory that began taking data on April 1, 1996, the solar
heliospheric observatory (SOHO) that was launched on December 2, 1995 and
is now observing the sun non stop, and the ground based telescopes in the
Global Oscillations Network Group (GONG) which have just begun observing
solar oscillations around the clock (for general reviews see Science, 31 
May 1996), may very soon point at the correct solution to the SNP. 

\section{Is There a $^8$B Solar Neutrino Problem?} 
Table I presents a comparison between the solar neutrino observations and
the SSM predictions of Bahcall and Pinsonneault 1995 (BP95) and of Dar and
Shaviv 1996 (DS96). Although BP (and some other similar SSM calculations)
predict a $^8$B solar neutrino flux that is approximately 2.6 larger than
observed by Super-Kamiokande, $\phi_{\nu_e}=2.51\pm 0.14(stat.)\pm 
0.18(syst.) ~\times 10^6~cm^{-2}s^{-1}$, (Suzuki 1997), DS predict a flux
consistent with that observed by Super-Kamiokande. The differences between BP
and DS are summarized in Table II (for details see Dar and Shaviv 1996). 
The difference between the predicted $^8$B flux are mainly due to the use
of updated nuclear reaction rates by DS, differences in the calculated
effects of diffusion, differences in the initial solar composition assumed
in the two calculations and the use of updated equation of state by DS.  
They reduce the predicted $^8$B flux
approximately by factors of 0.55, 0.81, 0.95 and 0.96, respectively (the
remaining differences are mainly due to inclusion of partial ionization
effects, premain sequence evolution and deviations from complete nuclear
equilibrium by DS which were neglected by BP, and due to different
numerical methods, fine zoning and time steps used in the two
calculations): 

\noindent {\bf Nuclear Reaction Rates:} 

The uncertainties in the nuclear reaction rates at solar conditions are
still large due to (1) uncertainties in the measured cross sections at
laboratory energies, (2) uncertainties in their extrapolations to solar
energies, (3) uncertainties in dense plasma effects (screening,
correlations and fluctuations) on reaction rates.  Rather than averaging
measured cross sections that differ by many standard deviations, DS used
for the extrapolation only the most recent and consistent measurements of
the relevant nuclear cross sections. Because sub-Coulomb reactions take
place when the colliding nuclei are far apart, the Optical Model and the
Distorted Wave Born Approximation give a reliable description of their
energy dependence.  DS have used them for extrapolating measured
sub-Coulomb cross sections to solar energies.  BP preferred to rely on
published extrapolations of averaged cross sections based on energy
dependences calculated from microscopic nuclear reaction models (e.g.
Johnson et al 1992). Similar screening corrections (which by accidental
cancellation have a very small net effect on $\phi_{\nu_\odot}(^8{\rm
B})$) have been used by DS and BP. The updated ``astrophysical S factors''
which were used by DS are listed in Table II. They reduce the BP
predictions by approximately a factor of 0.55. 

\noindent

{\bf Diffusion:} Diffusion, caused by density, temperature, pressure,
chemical composition and gravitational potential gradients play an
important role in the sun since it modifies the local chemical composition
in the sun. The relative changes in SSM predictions due to diffusion of
all elements are summarized in Table III. While BP found a rather large
increases in the predicted $^7$Be, $^8$B, $^{13}$N, $^{15}$O and $^{17}$F
solar neutrino fluxes; 14\%, 36\%, 52\%, 58\%, and 61\% which result in
36\%, 33\%, 9\% increases in their predicted rates in Super-Kamiokande,
Homestake, and in GALLEX and SAGE, respectively, DS found only a moderate
increase due to diffusion, 4\%, 10\%, 23\%, 24\% and 25\%, respectively,
in the above fluxes, which result in 10\%, 10\% and 2\% increase in the
predicted rates in Super-Kamiokande, Homestake, and in GALLEX and SAGE,
respectively. Although the two diffusion calculations assumed a different
initial solar chemical composition (see below) and BP approximated the
diffusion of all elements heavier than $^4$He by that of fully ionized
iron (the DS calculations followed the diffusion of each element
separately and used diffusion coefficients calculated for the actual
ionization state of each element at each shell in the sun as obtained from
solving the local Saha equations), these cannot fully explain the above
large differences.  Recent independent diffusion calculations by Richard
et al. (1996) obtained similar results to those obtained by DS as can be
seen from Table III (we interpolated the results from the two models of
Richard et al. to the initial chemical composition assumed by DS). 

\noindent 
{\bf Initial Chemical Composition:} The initial chemical composition
influences significantly the solar evolution and the present density,
chemical composition and temperature in the solar core, which determine
the solar neutrino fluxes.  In particular, the calculated radiative
opacities, which in turn determine the temperature gradient in the solar
interior, are very sensitive to the heavy elements abundances (the heavy
elements are not completely ionized in the sun).  Apart from the noble
gases, only a few elements such as H, C, N and O, which were
able to form highly volatile molecules or compounds,  have escaped complete
condensation in primitive early solar system meteorites (see, e.g., Sturenburg
and Holweger 1990).  Thus, the initial solar abundances of all other
elements are expected to be approximately equal to those found in type I
carbonaceous chondrites as a result of their complete condensation in the
early solar system.  Since the chemical composition of the solar surface
is believed to have changed only slightly during the solar evolution (by
nuclear reactions during the Hayashi phase, by diffusion and turbulent
mixing in the convective layer during the main sequence evolution, and by
cosmic ray interactions at the solar surface) it has been expected that
the photospheric abundances of these elements are approximately equal to
those found in CI chondrites.  Over the past decades there have been many
initial disagreements between the meteoritic and photospheric abundances.
In nearly all cases, when the atomic data were steadily improved and the
more precise measurements were made, the photospheric values approached
the meteoritic values. The photospheric abundances are now as a rule in
very good agreement with the meteoritic values (Grevesse and Noels 1991;
1993). 
Since the  meteoritic values represent the initial values and are known
with much better accuracy (often better than 10\%) than the photospheric
ones, DS assumed that the initial solar heavy metal abundances are given
approximately by the meteoritic (CI chondrites) values of Grevesse and
Noels (1993) and adjusted the initial CNO and Ne abundances to reproduce
their observed photospheric abundances.  Also the unknown initial $^4$He
solar abundance has been treated as an adjustable parameter.  DS
``predicted'' its present photospheric mass fraction to be $ Y=0.238\pm
0.05$ in good agreement with the $^4$He surface mass fraction inferred
from helioseismology:  $Y_s=0.242\pm 0.003$ (Hernandez and
Christensen-Dalsgaard 1994). However, their formal error is highly
misleading because of the great sensitivity of the result to the model of
the solar atmosphere, the equation of state there and the atmospheric
opacities. We estimate that at present the $^4$He mass fraction at the
solar surface is not known from helioseismology better than $Y_s=0.242\pm
0.010$).. BP adjusted the initial solar composition to reproduce 
the present day surface abundances which, except for the CNO and the nobel
gases, were assumed to be represented by their meteoritic values. 

\noindent 
The photospheric abundances of $^7$Li, $^9$Be and $^{11}$B are smaller by
a factor of nearly 150, 3 and 10, respectively, than their meteoritic
abundances. The origin of such large differences is still not clear.  They
cannot be explained by nuclear burning during the Hayashi phase although
significant Lithium burning does takes place during this phase. They may
be explained by rotational mixing (e.g., Richard et al 1996). Although the
initial solar (meteoritic) abundances of Lithium, Beryllium and Boron are
very small and do not play any significant role in solar evolution their
depletion perhaps can provide a clue to the real history of the convection
zone and the sun. 

\noindent
{\bf Equation of State:} 

The equation of state is used to calculate the local density and
temperature required to balance the gravitational pressure in the sun.
Since the neutrino producing reactions in the sun depend strongly on
temperature, their predicted fluxes depend strongly on the equation of
state. DS have used an updated equation of state which is described in
detail in DS96. It is consistent with the new OPAL equation of state
(Iglesias amd Rogers 1996). The use of an improved equation of state
reduce significantly our 1994 solar neutrino fluxes and improves the
agreement between the sound speed in the solar core that we calculated
from our SSM and the sound speed that is extracted from helioseismology.
The agreement with the updated sound speed from helioseismology
(Christensen Dalsgaard, 1996) is better than $2\times 10^{-3}$, as is
demonstrated in Fig. 1. It is significantly better than the agreement
obtain/reported by other SSM calculations, but shares, with all other
standard solar models (e.g., Bahcall et al. 1997), systematic deviations
from the measured sound speed. Note also that helioseismology has
confirmed only that the SSM describes quite well the ratio of pressure to
density in the present sun but it does not provide any evidence that the
production of $^7$Be, $^8$B and CNO solar neutrino fluxes are well
described by the SSM. 

\section{Neutrino Properties Beyond the Minimal SEM?}

Counting rates in $\nu_\odot$ experiments are formally given by
\begin{equation}
R=N_A\Sigma_i \phi_{\nu_\odot}(i)\int_{E_0}(dn_{\nu_i}/dE)\sigma_{\nu A}
(E)dE 
\end{equation}
where $N_A$ is the number of ``active'' atoms in the detector,
$\sigma_{\nu A}(E)$ is their cross section for neutrinos with energy E,
$dn_{\nu_i}/dE$ is the normalized energy spectrum of neutrinos
from reaction $i$ in the sun and $\phi_{\nu\odot}$ is their total flux.
Both, $dn_{\nu_i}/dE$ and $\sigma_{\nu A}$ follow directly from
the standard electroweak theory and are independent of the sun. 
($dn_{\nu_i}/dE$ is practically the standard $\beta$-decay
spectrum for the $\beta$-decays 2p$\rightarrow De^+\nu_e$,
$^8$B$\rightarrow 2\alpha e^+\nu_e$, $^{13}$N$\rightarrow
^{13}$C$e^+\nu_e$ and $^{15}$O$\rightarrow^{15}$N$e^+\nu_e$ and is a 
$\delta$-function for the electron captures 
$e^7$Be$\rightarrow\nu_e^7$Li
and $pep\rightarrow D\nu_e$.) Thus {\it conclusive evidence}
for new electroweak physics can be provided only by detecting
at least one of the following signals:

1. Spectral distortion of the SEM $\beta$-decay spectrum.

2. Solar neutrino flavors other than $\nu_e$.         

3. Terrestrial Modulations of solar neutrino fluxes.

4. A violation of the luminosity sum rule.

5. Rates which require negative $\phi_{\nu_\odot(i)}$. 

\noindent
So far, no such  clear evidence has been provided by the $\nu_\odot$
experiments. 

\noindent
{\bf Spectral Distortions:} At present only Super-Kamiokande can test
whether the spectrum of their detected $\nu_\odot$'s is consistent with
the $\nu_e$ spectrum from $\beta$-decay of $^8$B. So far Super-Kamiokande 
(and Kamiokande before) has observed (Suzuki 1997) an
electron recoil spectrum from $\nu_\odot e$ interactions which is
consistent, with that expected from an
undistorted $^8$B solar neutrino spectrum. Super-Kamiokande, which has been
running since April 1, 1996, will soon have  much more statistics
allowing a more sensitive test. 

\noindent {\bf Flavour and/or Helicity Flip}: 
Neutrino oscillations or neutrino helicity flip can explain the solar
neutrino observations (see, e.g., Voloshin these proceedings). However, no
time variation which is predicted by a magnetic helicity flip has been
detected by the $\nu_\odot$ experiments. The present solar neutrino
experiments can neither detect (Homestake, GALLEX and SAGE) nor
distinguish (Super-Kamiokande) between different neutrino flavors. However,
Super-Kamiokande will soon be able to examine with a high level of
sensitivity (real time, high statistics) whether the $^8$B solar neutrino
flux is time dependent. The sensitivity of Super-Kamiokande to
temporal variation in the solar neutrino flux will be demonstrated by
measuring the annual variation of the flux due to the annual variation of
the distance of Earth from the sun. Any other confirmed variations which 
depend
on the distance of Earth from the sun, on the orientation of the Earth
relative to the sun (summer-winter) and on the  
(day-night, summer-winter) or distance dependence will signal 
with day 
Only future experiments like SNO will 
be able to detect other neutrino flavors. 

\noindent {\bf The Solar Luminosity Sum Rule:} If the sun derives its
energy from fusion of Hydrogen into Helium and {\it if it is in a steady
state} where its nuclear energy production rate equals its luminosity,
then conservation of baryon number, electric charge, lepton flavor and
energy requires that the total solar neutrino flux at Earth satisfies
(e.g., Dar and Nussinov 1991): \begin{equation}
\phi_{\nu_\odot}={2L_\odot\over Q-2\bar{E}_\nu}~{1\over 4\pi D^{2}} \geq
6.52\times 10^{10}~cm^{-2} s^{-1}~, \end{equation} where $D\approx 1.496
\times 10^{13}~cm$ is the distance to the sun, $Q=26.733~MeV$ is the
energy released when four protons fuse into Helium,
$\bar{E}_\nu=\sum{E_{\nu_i}}\phi_{\nu_i }/ \sum{\phi_{\nu_i}}$ is the
average energy of solar neutrinos and $\bar{E}_\nu\geq 0.265~MeV$ if the
pp reaction in the sun produces $\nu_\odot$'s with the smallest average
energy. Eq. (2) can be rewritten as a luminosity sum rule: 
\begin{equation} \Sigma_i(Q/2-\bar{E}_{\nu_i})\phi_{\nu_i}= S,
\end{equation} where $S=L_\odot/ 4\pi D^2=1367~W~m^{-2}$ is the solar
``constant''.  A clear Violation of eq. (2) or the solar luminosity sum
rule, can prove that lepton flavor is not conserved. The Gallium
experiments with the low energy threshold of 233 keV, which makes them
sensitive to almost all the SSM neutrinos, reported updated time-averaged
capture rates of $70\pm 8~SNU $ in GALLEX and $72\pm 12~ SNU$ in SAGE
(see, e.g., Gavrin these proceedings).  These rates, smaller than that
predicted by SSMs are still consistent within the experimental
uncertainties with $76\pm 2~SNU$, the ``minimal'' signal expected from eq.
(2) if $\sigma_{Ga}= (1.18\pm 0.02)\times 10^{45}~cm^{-2}$, and all the
$\nu_\odot$'s were pp $\nu$'s.  However, the $^8$B solar neutrino flux
measured in Super-Kamiokande, $\phi_{\nu_\odot}=(2.51\pm 0.4)\times
10^6~cm^{-2}$, contributes another $6\pm 2~SNU$ which increase the minimal
expected signal in Gallium to $82\pm 3~$ SNU.  This somewhat larger rate
is still consistent within $2\sigma$ with the capture rates measured by
GALLEX and SAGE, in particular if their rates are ``recalibrated'' using
their Cr source experiments (Gavrin, these proceedings). However, the
Gallium experiments leave no room for significant (SSM-like) contributions
from $^7$Be and CNO solar neutrinos. This confirms the combined results
from the Chlorine experiment at Homestake and from the Kamiokande and
Super-Kamiokande experiments which seem to indicate that
$\phi_{\nu_\odot}(^8$B) is strongly supressed (see below).

\noindent
{\bf Missing $\nu_\odot$'s in $^{37}$Cl ?}: Althogh the $^{37}$Cl experiment 
with an energy threshold of $814~keV$ is completely blind to the pp solar
neutrinos it is sensitive to both the $^8$B neutrinos and the lower energy
$pep$, CNO and $^7$Be neutrinos. However, while the expected signal from a
$^8$B solar neutrino flux alone as measured by Super-Kamiokande is $2.75\pm
0.28~SNU$, the time-averaged counting rate in the $^{37}$Cl experiment is
$2.56\pm 0.25~SNU$ (see, e.g., Gavrin, these proceedings).  Although the
$^{37}$Cl experiment has not been ``calibrated'' with a neutrino source,
the Cr source experiments of GALLEX and SAGE suggest that the accuracy of
the radiochemical experiments is probably of the order of 10\%, or better.
Consequently, although the joint results from Homestake and Kamiokande do
not provide solid evidence for ``new electroweak physics'' (e.g., Bahcall
and Bethe 1991) they indicate that the combined contributions from
$^7$Be, CNO and pep solar neutrinos is strongly suppressed in $^{37}$Cl
compared with their SSM estimated contribution. 

\section{Are $^7$Be Solar Neutrinos Missing?} 
Electron capture by $^7$Be into the ground state of $^7$Li produces 862
keV neutrinos. The threshold energy for neutrino absorption by $^{37}$Cl
is 814 keV. Thus, absorption of $^7$Be neutrinos by $^{37}$Cl produces 48
keV electrons.  The maximum energy of the pp solar neutrinos is 420 keV.
The threshold energy for neutrino absorption in $^{71}$Ga $(3/2^-)$ is 233
keV into the ground state $(1/2^-)$ and 408 into its first excited state
$(5/2^-)$. The produced electrons have therefore energies below 187
and 12 $keV$, respectively. If the theoretical cross sections for neutrino
absorption near threshold overestimate significantly their true values
then the predicted rates will significantly overestimate the expected
signals in the Chlorine and Gallium experiments.
\noindent
An indication that final state interactions effects are not completely
understood is provided by Tritium $\beta$-decay.  Although final state
interactions in Tritium $\beta$-decay have been studied extensively, they
do not explain well the end-point $\beta$-decay spectrum ($E_e\sim
18.6~keV$). In all recent measurements, the measured spectrum yields a
negative value for the fitted squared mass of the electron neutrino. 
Final state interactions effects (screening of the nuclear charge by
atomic electrons, exchange effects, radiative corrections, nuclear recoil
against the electronic cloud, etc) in neutrino captures near threshold in
$^{37}$Cl and $^{71}$Ga may be much larger because their Z values are much
larger and because the de Broglie wave lengths of the produced electrons
are comparable to the Bohr radii of the atomic K shells in Cl and Ga. If
final state interactions reduce considerably the near threshold absorption
cross sections of pp neutrinos in $^{71}$Ga (making room for the expected
contribution of $^7$Be solar neutrinos in Gallium) and of $^7$Be neutrinos
in $^{37}$Cl, perhaps they can make the solar neutrino observations of
Super-Kamiokande and the Homestake experiments compatible.  Such a
terrestrial solution of the solar neutrino problem implies that
experiments such as BOREXINO and HELLAZ will observe the full $^7$Be solar
neutrino flux. 

\section{Astrophysical Solutions To The SNP}
\bigskip
Even if the $^7$Be solar neutrino flux is strongly suppressed, it does not
eliminate standard physics solutions to the solar neutrino problem: 
\noindent
The ratio between the fluxes of $^7$Be and $^8$B solar neutrinos 
is given by  
\begin{equation} 
R={\phi_{\nu_\odot}(^7{\rm Be})\over \phi_{\nu_\odot}(^8{\rm B})}
 = {\int n_e n_7<\sigma v>_{e7}4\pi r^2dr\over
   \int n_p n_7<\sigma v>_{p7}4\pi r^2dr}.
\end{equation}
Because of the decreasing temperature and Be7 abundance as function of
distance from the center of the sun on the one hand, and the $\sim r^2$ 
increase in radial mass on
the other, the production of $^7$Be and $^8$B solar neutrinos
in the SSM peaks around an effective radius, $r_{eff}\approx
0.064R_\odot$ ($r_{eff}$ is approximately the radius within which 50\% of
the flux is produced) . The SSM also predicts a ratio of electron to
proton densities near the center of the sun, $n_e/n_p\sim 2$, consistent
with helioseismology observations.  Consequently, the SSMs predict  
\begin{equation}
R\approx {2<\sigma v>_{e7}\over <\sigma v>_{p7}} \approx 1.27\times 
10^{-14}S_{17}^{-1}F_{17}^{-1}T_7^{1/6}e^{47.625/T_7^{1/3}},
\end{equation}      
where $F_{17}$ is the screening correction to the p-capture rate by $^7$Be,  
$T_7$ is the temperature in $10^7K$ at the effective radius and   
$S_{17}$ is in $eV~barn$ units.
The SSMs yield $T_7(r_{eff})\approx 1.45$. Using $S_{17}(0)=17~eV~b$  
and  $\phi_\odot(^8{\rm B})=2.51\times 10^6~cm^{-2}~s^{-1}$ as observed
by Super-Kamiokande , one can reproduce the SSM prediction (e.g., Dar and 
Shaviv 1996) 
\begin{equation}
\phi_{\nu_\odot}(^7{\rm Be})=R 
\phi_{\nu_\odot}(^8{\rm B})\approx 3.7\times 10^9~cm^{-2}~s^{-1}. 
\end{equation}
Astrophysical solutions of the solar neutrino problem aim towards
suppressing the value of R. Three alternatives are currently investigated:
 
\noindent 
{\bf Plasma Physics Effects}: The effects of the surrounding
plasma on nuclear reaction rates in dense stellar plasmas, and in
particular on proton and electron capture by $^7$Be in the sun are known
only approximately. Because of accidental cancellations the screening
corrections in the Debye approximation (Salpeter and Van Horn 1969) to the
rates of all nuclear reactions do not change the predicted $^8$B solar
neutrino flux, although the screening corrections to the individual
reactions are considerable (an enhancement by a factor $F\approx
e^{Z_1Z_2e^2/kTr_{D}}$ where $r_{D}$ is the Debye screening radius. The
conditionns required for the applicability of of the Debye approximation
are not satisfied in the solar core. In particular, because of the small
number of particles within the Debye sphere, $n\sim 1$, plasma
fluctuations may affect significantly the nuclear reaction rates. It is
interesting to note that screening have the opposite effect on e-capture
(reduction) compared with p capture (enhancement). In order to explain the
deficit of $^7$Be solar neutrinos, plasma screening effects must enhance
considerably the ratio between electron and proton capture by $^7$Be,
relative to that predicted by the weak screening theory (Salpeter and Van
Horne 1969).  Perhaps a more exact treatment of screening, may change R
considerably.  This possibility is currently studied, e.g., by Shaviv and
Shaviv (1996) using numerical methods and by Brown and Sawyer (1996) using
quantum statistical mechanics techniques. 

\noindent 
In principle, collective plasma physics effects, such as very strong
magnetic or electric fields near the center of the sun, may polarize the
plasma electrons, and affect the branching ratios of electron capture by
$^7$Be (spin $3/2^-$) into the ground state (spin $3/2^-$,
$E_{\nu_e}=0.863~MeV$, BR=90\%) and the excited state (spin $1/2^-$,
$E_{\nu_e}=0.381~MeV$, BR=10\%) of $^7$Li. Since solar neutrinos with
$E_{\nu_e}=0.381~MeV$ are below the threshold (0.81 MeV) for capture in
$^{37}$Cl and have a capture cross section in $^{71}$Ga that is smaller by
about a factor of 6 relative to solar neutrinos with
$E_{\nu_e}=0.863~MeV$, therefore a large suppression in the branching
ratio to the ground state can produce large suppressions of the $^7$Be
solar neutrino signals in $^{37}$Cl and in $^{71}$Ga. However, such an
explanation requires anomalously large fields near the center of the sun. 
 
\noindent
{\bf Temporal and Spatial Variations in T:}
Davis (1996) has been claiming persistently that the solar neutrino flux
measured by him and his collaborators in the $^{37}$Cl radiochemical
experiment is varying with time. Because of the possibility that neutrinos
may have anomalous magnetic moments, much larger than those predicted by
minimal extensions of the standard electroweak model, which can solve the
solar neutrino problem, attention has
been focused on anticorrelation between the solar magnetic activity (the
11 year cycle) and the $\nu_\odot$ flux (see, e.g., Davis 1996). Also a
day-night effect (e.g., Cribier et al 1986; Dar and Mann 1987) due to
resonant conversion of the lepton flavor of solar neutrinos which cross
Earth at night before reaching the solar neutrino detector was not found
by Super-Kamiokande. However, the basic general question whether the solar
neutrino flux varies on a short time scale has not been fully answered
by the first generation of solar neutrino experiments, mainly because of 
limited statistics.
\noindent
The SSM predicts that there is no significant variation of the solar 
neutrino  flux on time scales shorter than millions of years. However, the
sun has a
differential rotation. It rotates once in $\sim$ 25 days near the equator,
and in $\sim$ 33 days near the poles. Moreover, the observed surface
rotation rates of young solar-type stars are up to 50 times that of the
sun. It suggest that the sun has been loosing angular momentum over its
lifetime. The overall spin-down of a sun-like star by mass loss and
electromagnetic radiation is difficult to estimate from stellar evolution
theory, because it depends on delicate balance between circulations and
instabilities that tend to mix the interior and magnetic fields that
retard or modify such processes. It is quite possible that the
differential rotation extends deep into the core of the sun and causes
there spatial and temporal variations in the solar properties due to
circulation, turbulences and mixing. Since R is very sensitive to the
temperature, even small variations in temperature can affect R
significantly without affecting significantly the pp solar neutrino flux
(the $^8$B solar neutrinos will come mainly from temperature
peaks, while the pp neutrinos will reflect more the average temperature). 
Another possibility is that the g-modes in the convective layer induce
temporal and spatial variations in the solar core with a significant
amplitude. 
\noindent
In fact, a cross correlation analysis of the various data sets from the
Homestake, Kamiokande, GALLEX and SAGE, shows an unexpected correlation:
If arbitrary time lags are added to the different solar neutrino
experiments, the cross correlation is maximal when these time lags vanish. 
Moreover, a power spectrum analysis of the signals shows a peak around 21
days, suggesting a periodical variation (Sturrock and Walther 1996). The
effect may be a statistical fluke. However, it can also indicate a real
short time scale variation in the solar core. Fortunately, Super-Kamiokande
will soon provide the answer to whether the $^8$B solar neutrino flux is
time-dependent or not (We propose to do a similar cross correlation 
analysis between the observed rates in different halves of the detector 
in order to see whether variations in the total counting 
rate, other than due to the eccentricity of the orbit of Earth around 
the sun, are due to statistical Poisson noise or reflect a time
dependent signal).
\noindent 
{\bf Mixing of $^3$He:} 
The SSM $^3$He equilibrium abundance increases sharply with radius.
Cumming and Haxton (1996) have recently suggested that the $^7$Be solar
neutrino problem could be circumvented in models where $^3$He is
transported into the core in a mixing pattern involving rapid filamental
flow downward. We note that if this mixing produces hot spots (due to
enhanced energy release) they can increase the effective temperature for p
capture by $^7$Be in a cool environment, reducing R while keeping the
$^8$B solar neutrino flux at the observed level. Perhaps, helioseismology
will be able to test that. 

\noindent
Cummings and Haxton (1996) also noted that such mixing will have other
astrophysical consequences. For example, galactic evolution models predict
$^3$He abundances in the presolar nebula and in the present interstellar
medium (ISM) that are substantially (i.e., a factor of five or more) in
excess of the observationally inferred values.  This enrichment of the ISM
is driven by low-mass stars in the red giant phase, when the convective
envelope reaches a sufficient depth to mix the $^3$He peak, established
during the main sequence, over the outer portions of the star.  The $^3$He
is then carried into the ISM by the red giant wind. The core mixing lowers
the main sequence $^3$He abundance at large r. 
 
\section {The MSW Solution}
Standard solar models, like the one calculated by Dar and Shaviv (1996),
perhaps can explain the results reported by Kamiokande. However, if the
neutrino absorption cross sections used by the radiochemical experiments
are correct, then standard physics cannot explain an $^{37}$Ar production
rate in $^{37}$Cl smaller than that expected from the solar $^8$B neutrino
flux measured by Kamiokande (assuming that both results are correct). If
the experimental results of Kamiokande and Homestake are interpreted as an
evidence for such a situation (e.g., Bahcall 1994; 1995), they do imply
new physics beyond the standard particle physics model (Bahcall and Bethe
1991). In that case an elegant solution to the solar neutrino anomaly is
resonant neutrino flavor conversion in the sun, first proposed by Mikheyev
and Smirnov (1986) (see also Wolfenstein 1978; 1979). It requires only a
natural extension of the minimal standard electroweak theory and it is
based on a simple quantum mechanical effect. Many authors have carried out
extensive calculations to determine the neutrino mixing parameters which
can bridge between the predictions of the standard solar models and the
solar neutrino observations. They found that a neutrino mass difference
$\Delta m^2\sim 0.7\times 10^{-5}~eV^2 $ and a neutrino mixing of $sin^2
2\theta\approx 0.5\times 10^{-2}$ can solve the solar neutrino problem
(see, e.g., Gavrin, these proceedings). These parameters, however, cannot 
explain the neutrino-oscillation-like signal which was reported by the LSND
experiment (Athanassopoulos 1996).

\section{Conclusions} 
The solar neutrino problem may be an astrophysical problem. An indication
for that may come from observation of unexpected temporal variability of
the solar neutrino flux by Super-Kamiokande or from helioseismology
observations by SOHO and GONG.  An indication may also come from cross
correlation analysis of the time dependent of the counting rates in GALLEX
and Sage and of the counting rates of Kamiokande and Homestake. Such cross
correlation analysis may test whether the time variation of the counting
rates is statistical or physical. Deviations of the experimental results
from those predicted by the standard solar models may reflect the
approximate nature of these models (which neglect angular momentum effects,
differential rotation, magnetic field, angular momentum loss and mass loss
during evolution and do not explain yet, e.g., solar activity and the
surface depletion of Lithium, Berilium and Boron relative to their
meteoritic values, that may or may not be relevant to the solar neutrino
problem). Improvements of the standard solar model should continue. 
In particular, dense plasma effects on nuclear reaction rates and radiative
opacities, which are not well understood, may affect the SSM predictions
and should be further studied, both theoretically and experimentally.
Relevant information may be obtained from studies of thermonuclear plasmas
in inertial confinement experiments. Useful information may also be
obtained from improved data on screening effects in low energy nuclear
cross sections of ions, atomic beams and molecular beams incident on a
variety of gas, solid and plasma targets. 
 
\noindent
Better knowledge of low energy nuclear cross sections is badly needed.
Measurement of crucial low energy nuclear cross sections by new methods,
such as measurements of the cross sections for the radiative captures
${\rm p+^7Be\rightarrow ^8B+\gamma}$ and ${\rm ^3He+^4He\rightarrow
^7Be+\gamma}$ by photodissociation of $^8$B and $^7$Be in the coulomb
field of heavy nuclei are badly needed in order to determine whether there
is a $^8$B solar neutrino problem. 
 
\noindent
The $^{37}$Ar production rate in $^{37}$Cl indeed may be smaller than that
expected from the flux of standard solar neutrinos as measured by electron
scattering in the Kamiokande experiment. In that case neutrino
oscillations, and in particular the MSW effect, may be the correct
solution to the solar neutrino problem. Only future experiments, such as
SNO, Super-Kamiokande, BOREXINO and HELLAZ, will be able to supply a
definite proof that Nature has made use of this beautiful effect. 

\noindent
{\bf Acknowledgements:} The author would like to thank the organizers of
the ITEP 1997 International Winter School on Particle Physics for the
invitation and the warm hospitality extended to him.  This talk is based
on an ongoing collaboration with Giora Shaviv which is supported in part
by the Technion fund for the promotion of research.

\clearpage 
 
\noindent
{\bf Table Ia:} Comparison between the solar neutrino fluxes predicted
by the SSM of BP95 and of  
DS96, and measured by the four solar neutrino experiments.
$$\matrix{~~~~~\nu~{\rm Flux}\hfill & {\rm BP95} \hfill & {\rm   DS96}\hfill 
&{\rm  Observations} \hfill &{\rm Experiment} \hfill \cr
{\rm \phi }_{\nu }(pp)~[{10}^{10}{cm}^{-2}{s}^{-1}] \hfill &5.91
\hfill &6.10 \hfill  &\mit &\rm \cr
{\phi }_{\nu }(pep)~[{10}^{8}{cm}^{-2}{s}^{-1}] \hfill &
1.39\hfill &1.43\hfill &\rm \hfill &\rm \cr
{\phi }_{\nu }(^{7}{Be})~[{10}^{9}{cm}^{-2}{s}^{-1}]\hfill &5.18\hfill
&3.71\hfill &\rm \hfill
&\rm \hfill \cr
{\phi }_{\nu }(^{8}{B})~[{10}^{6}{cm}^{-2}{s}^{-1}]\hfill &6.48\hfill
& 2.49 \hfill &\rm 2.51\pm 0.23\hfill &\rm Super-Kamiokande\cr
{\phi }_{\nu }(^{13}{N})~[{10}^{8}{cm}^{-2}{s}^{-1}]\hfill &6.4\hfill
& 3.82 \hfill &\rm \hfill &\rm \cr
{\phi }_{\nu }(^{15}{O})~[{10}^{8}{cm}^{-2}{s}^{-1}]\hfill &5.15\hfill
& 3.74 \hfill &\rm \hfill &\rm \cr
{\phi }_{\nu }(^{17}{F})~[{10}^{6}{cm}^{-2}{s}^{-1}]\hfill &6.48\hfill
& 4.53 \hfill &\rm \hfill &\rm \cr
\hfill& \hfill & \hfill &\hfill\cr
\Sigma (\phi \sigma)_{Cl}~[SNU]\hfill&  9.3\pm 1.4\hfill &
4.1\pm 1.2\hfill & 2.56\pm 0.25
 \hfill &{\rm Homestake}\hfill \cr
\Sigma(\phi\sigma)_{Ga}~[SNU]\hfill & 137\pm 8\hfill
& 115\pm 6\hfill
&70\pm 8 \hfill &{\rm GALLEX}\cr
\Sigma(\phi\sigma)_{Ga}~[SNU]\hfill & 137\pm 8\hfill
& 115\pm 6\hfill
&72\pm 12 \hfill &{\rm SAGE}\cr}$$

\noindent
{\bf Table Ib} Characteristics of the BP95, DS94, and DS96
Solar Models in Table Ia
(c=center; s=surface; bc=base of convective zone;
${\rm \bar N=log([N]/[H])+12)}$.
 
$$\matrix{{\rm Parameter}\hfill& {\rm BP95}\hfill& {\rm
DS94}\hfill& {\rm
DS96}\hfill \cr
{T}_{c}~[{10}^{7}K] \hfill &1.584 \hfill &1.554 \hfill &1.561 \hfill \cr
{\rho }_{c}~[g~c{m}^{-3}]\hfill&156.2 \hfill&155.3\hfill&155.4\hfill
\cr
{X}_{c}\hfill&0.3333\hfill &0.3462\hfill &0.3424 \hfill \cr
{Y}_{c}\hfill&0.6456 \hfill &0.6359 \hfill &0.6380 \hfill \cr
{Z}_{c}\hfill&0.0211\hfill&0.01950 \hfill
&0.01940 \hfill \cr
{R}_{conv}~[R/R_{\odot}]\hfill&0.712 \hfill &0.7105 \hfill &0.7130
\hfill \cr
{T}_{bc}~[{10}^{6}{\rm K}]\hfill&2.20 \hfill &2.10 \hfill &2.105
\hfill \cr
{X}_{s}\hfill&0.7351 \hfill &0.7243 \hfill &0.7512 \hfill \cr
{Y}_{s}\hfill&0.2470 \hfill &0.2597 \hfill &0.2308 \hfill \cr
{Z}_{s}\hfill&0.01798 \hfill &0.01574 \hfill
&0.0170 \hfill \cr
\overline{N}_s{(^{12}C})\hfill&8.55\hfill &8.50 \hfill&8.55 \hfill \cr
\overline{N}_s{(^{14}N})\hfill&7.97 \hfill &7.92 \hfill&7.97\hfill \cr
\overline{N}_s{(^{16}O})\hfill&8.87 \hfill &8.82 \hfill&8.87 \hfill \cr
\overline{N}_s{(^{20}Ne})\hfill&8.08 \hfill &8.03 \hfill&8.08 \hfill
\cr
{T}_{eff}~[{\rm K}]\hfill& \hfill &5920 \hfill &5803
\hfill \cr}$$
 
\clearpage 
 
\noindent
{\bf Table II:} Comparison between the SSM of Bahcall and Pinsonneult
(1995) and of Dar and Shaviv (1996).
$$\matrix{  
\hfill& {\rm ~~~~~~BP95}\hfill& {\rm 
~~~~~~DS96}\hfill \cr
\hfill&\hfill& \hfill \cr
{M}_{\odot} \hfill &1.9899\times 10^{33}~g \hfill &1.9899\times 10^{33}~g 
\hfill\cr
{L}_{\odot} \hfill&3.844\times 10^{33}~erg~s^{-1} \hfill& 3.844\times 10^{33}
~erg~s^{-1}\hfill\cr
{R}_{\odot}\hfill&6.9599\times 10^{10}~cm \hfill &6.9599\times 
10^{10}~cm\hfill \cr 
{t}_{\odot}\hfill&4.566\times 10^9~ y \hfill &4.57\times 10^9~y 
\hfill \cr 
{\rm Rotation}\hfill&{\rm Not~Included} \hfill & {\rm Not~Included} 
\hfill \cr 
{\rm Magnetic~Field}\hfill&{\rm Not~Included} \hfill & {\rm Not~Included} 
\hfill \cr 
{\rm Mass ~Loss}\hfill&{\rm Not~Included} \hfill & {\rm Not~Included} 
\hfill \cr 
{\rm Angular~Momentum~Loss}\hfill&{\rm Not~Included} \hfill & {\rm Not 
~Included} \hfill \cr 
{\rm Premain~Sequence~Evolution}\hfill&{\rm Not~ 
Included} \hfill & {\rm Included} \hfill \cr 
{\rm Initial~Abundances: }\hfill& \hfill & \hfill \cr
{\rm ^4He}\hfill& {\rm Adjusted~Parameter} \hfill &{\rm 
Adjusted~Parameter} \hfill \cr 
{\rm C,N,O,Ne}\hfill& {\rm Adjusted~Photospheric} \hfill &{\rm 
Adjusted~Photospheric} \hfill \cr 
{\rm All~Other~Elements}\hfill& {\rm Adjusted~``Photospheric''} \hfill &{\rm 
Meteoritic} \hfill \cr 
{\rm Photospheric~Abundances: }\hfill&\hfill & \hfill \cr
{\rm ^4He }\hfill&{\rm Predicted} \hfill&{\rm Predicted}\hfill \cr
{\rm C,N,O,Ne}\hfill& {\rm Observed} \hfill &{\rm Observed} \hfill \cr 
{\rm All~Other~Elements}\hfill& {\rm =~Meteoritic} \hfill &{\rm 
Predicted} \hfill \cr 
\hfill& \hfill & \hfill \cr
{\rm Radiative~Opacities}\hfill&{\rm OPAL~1994}
\hfill & {\rm OPAL~1996} \hfill \cr 
{\rm Equation~ of~ State}\hfill&{\rm Straniero ~1988?}
\hfill & {\rm DS~ 1996} \hfill \cr 
{\rm Partial~ Ionization~ Effects}\hfill&{\rm Not~ 
Included} \hfill & {\rm Included} \hfill \cr 
{\rm Diffusion~ of~ Elements:}\hfill& \hfill & \hfill \cr
{\rm H,~^4He}\hfill&{\rm Included} \hfill&{\rm Included} \hfill \cr
{\rm Heavier~Elements}\hfill&{\rm Approximated~ by~ Fe } \hfill&{\rm 
 All~Included} \hfill \cr
{\rm Partial~ Ionization~ Effects}\hfill&{\rm Not~Included } \hfill&{\rm 
 Included} \hfill \cr
{\rm Nuclear~ Reaction~ Rates:}\hfill& \hfill & \hfill \cr 
S_{11}(0)\hfill &
\rm 3.896\times {10}^{-22}~keV\cdot b \hfill &\rm 4.07\times {10}^{-22}~
keV\cdot b \hfill\cr
S_{33}(0)\hfill &
4.99\times {10}^{3}~keV\cdot b \hfill &5.6\times {10}^{3}~keV\cdot b 
\hfill \cr S_{34}(0)\hfill &0.524~keV\cdot b \hfill & 
0.45~keV\cdot b\hfill \cr S_{17}(0)\hfill&0.0224~keV\cdot b \hfill
& 0.017~keV\cdot b\hfill \cr
{\rm Screening~ Effects}\hfill &{\rm Included } \hfill&{\rm Included} 
\hfill \cr 
{\rm Nuclear~Equilibrium}\hfill&{\rm Imposed} \hfill&{\rm Not~Assumed} 
\hfill \cr}$$
\clearpage

\clearpage
 
\noindent
{\bf Table III:} Fractional change in the predicted $\nu_\odot$ fluxes 
and counting rates in the $\nu_\odot$ experiments due to the inclusion of 
element diffusion in 
the SSM calculations of Bahcall and Pinsonneault (1996), Dar and Shaviv 
(1994, 1996) and Richard, Vauclair, Charbonnel and Dziembowski (1996).
The results of models 1 and 2 of RVCD were   
extrapolated to the initial solar composition which was used in DS96. 
 
$$\matrix{\phi_{\nu_\odot} & {\rm BP95}& {\rm DS96} & {\rm RVCD}\cr
pp\hfill &-~1.7\%\hfill &-~0.3\%\hfill &-~0.8\% \hfill\cr
pep\hfill &-~2.8\%\hfill &-~0.3\%\hfill &-~0.4\%\hfill\cr
{\rm ^7Be}\hfill    &+13.7\%\hfill &+4.2\%\hfill  &+ ~6.5\%\hfill\cr
{\rm ^8B}\hfill    &+36.5\%\hfill &+11.2\%\hfill &+10.7\%\hfill\cr
{\rm ^{13}N}\hfill &+51.8\%\hfill &+22.7\%\hfill &+19.8\%\hfill\cr
{\rm ^{15}O}\hfill &+58.0\%\hfill &+24.0\%\hfill &+20.8\%\hfill\cr
{\rm ^{17}F}\hfill &+61.2\%\hfill &+24.9\%\hfill &+21.8\%\hfill\cr
{\rm Rates} & & & {\rm RVCD}\cr
{\rm H2O}\hfill &+36.5\%\hfill &+11.2\%\hfill&+13.3\%\hfill\cr
{\rm Cl}\hfill &+32.9\%\hfill &+~9.5\%\hfill&+12.3\%\hfill\cr
{\rm Ga}\hfill &+~8.7\%\hfill &+~2.6\%\hfill&+~3.7\%\hfill
\cr}$$
 
\vfill
\eject
 
\end{document}